\documentclass[reprint,prx,superscriptaddress, amsmath,amssymb, aps]{revtex4-2}
\usepackage[dvipsnames]{xcolor}
\usepackage[english]{babel}
\usepackage{graphicx}
\usepackage{nccmath}
\usepackage{float}
\usepackage[utf8]{inputenc}
\usepackage[colorlinks,allcolors=cyan!70!black]{hyperref}
\usepackage{physics,amsmath,nccmath}
\usepackage{mathtools}
\usepackage{lipsum}
\newlength{\arrow}
\newlength{\extralongarrow}
\newcommand*{\myextralongrightarrow}[1]{\xrightarrow{\mathmakebox[\extralongarrow]{#1}}}

\newcommand{\Cov}{\mathrm{Cov}}

\newcommand{\Var}{\mathrm{Var}}

\newcommand{\cvratiofrac}{\frac{\textrm{CV}_\mathrm{\bar{x}}}{\textrm{CV}_\mathrm{\bar{y}}}}

\newcommand{\loglog}[2]{\frac{\partial \log #1}{\partial\log #2}}

\begin{document}
\title{Using random perturbations to infer the structure\\ of feedback control in gene expression}
\author{Seshu Iyengar}
\affiliation{Department of Physics, University of Toronto, 60 St.~George St., Ontario M5S 1A7, Canada}
\affiliation{Department of Chemical \& Physical Sciences, University of Toronto, Mississauga, Ontario L5L 1C6, Canada}
\author{Andreas Hilfinger}
\email[andreas.hilfinger@utoronto.ca]{}
\affiliation{Department of Physics, University of Toronto, 60 St.~George St., Ontario M5S 1A7, Canada}
\affiliation{Department of Chemical \& Physical Sciences, University of Toronto, Mississauga, Ontario L5L 1C6, Canada}
\affiliation{Department of Cell \& Systems Biology, University of Toronto, 25 Harbord St, Toronto, Ontario M5S 3G5}
\affiliation{Department of Mathematics, University of Toronto, 40 St.~George St., Toronto, Ontario M5S 2E4}

\begin{abstract}
Feedback in cellular processes is typically inferred through cellular responses to experimental perturbations. Modular response analysis provides a theoretical framework for translating specific perturbations 
into feedback sensitivities between cellular modules. However, in large-scale drug perturbation studies the effect of any given drug may not be known and 
may not only affect one module at a time. 
Here, we analyze 
the response of gene expression models to random perturbations that affect multiple modules simultaneously. In the deterministic regime we analytically show how cellular responses to infinitesimal random perturbations can be used to infer the nature of feedback regulation in gene expression, as long as the effects of perturbations are statistically independent between modules.
We numerically extend this deterministic analysis to the response of average abundances of stochastic gene expression models to finite perturbations. 
Across a large sample of stochastic models, the response of average abundances 
generally obeyed predicted bounds from the deterministic analysis, but 
dramatic deviations occurred in systems with bimodal or fat-tailed stationary state distributions. 
These discrepancies demonstrate 
how deterministic analyses can fail to capture the effect of perturbations on averages of stochastic cellular feedback systems---even in the linear response regime.
\end{abstract}
\maketitle

\section*{Introduction}
A mechanistic understanding of cellular processes requires characterizing how molecular abundances in cells affect each other. The response of molecular abundances to experimental perturbations, such as drug exposure, is a common starting point to infer molecular interactions. 
Modular response analysis (MRA) and related techniques~\cite{bruggeman2002modular,Andrec2005Inference,mochizuki2016} provide a general framework based on dynamical systems theory to translate experimental perturbation data into models of molecular feedback control. 

These deterministic approaches use a linear response assumption to calculate sensitivity coefficients of molecular abundances to perturbations in kinetic parameters of other network components~\cite{kholodenko2002untangling}. Sensitivities are then assumed to reflect the interactions between cellular components~\cite{santra2018reconstructing}. 
Crucially, MRA requires a fundamental restriction on perturbation targets: to infer the sensitivity of a particular component of interest to other cellular components, the perturbation must not directly perturb the kinetic parameters governing the dynamics of the component of interest~\cite{santra2018reconstructing}. 
Therefore, MRA requires some knowledge about the nature of experimental perturbations. However, this can be challenging to satisfy in biological high-throughput experiments that report hundreds of potentially uncharacterized drug perturbations~\cite{kaufman2022visual}.

Here, we analyze the response of gene regulatory models to random drug perturbations with multiple targets.
We find that the response of deterministic gene regulation models to random  perturbations is constrained by the feedback sensitivity of the mRNA's synthesis rate when perturbations are infinitesimal and affect protein reactions statistically independently from other cellular reactions.

Furthermore, we analyzed how average abundances in stochastic gene expression models respond to random perturbations to address whether deterministic methods can qualitatively infer feedback in intrinsically stochastic cellular processes. Previous research has shown that inferring feedback sensitivities from perturbations can be robust in the presence of measurement noise~\cite{kang2015discriminating,santra2018reconstructing,thomaseth2018impact,Klinger2018Reverse,borg2023modular,borg2024testing,Andrec2005Inference}. 
Additionally, MRA techniques have been tested on data where approximated stochastic noise is added to deterministic dynamical systems~\cite{carre2017reverse,marbach2010revealing,schaffter2011genenetweaver}.
While some prior MRA work has claimed to account for aspects of cellular stochasticity~\cite{santra2018reconstructing},
the effect of the intrinsically stochastic nature of biochemical reactions on cellular perturbation responses has not been explicitly analyzed within the framework of MRA.

We find that across a large sample of stochastic gene expression models with nonlinear feedback, numerical responses to finite perturbations generally obeyed the deterministic constraints. However, we also observed
that the response of average abundances in stochastic biochemical reaction networks can dramatically violate these constraints.
These discrepancies would lead to qualitatively incorrect conclusions from deterministic methods, and typically occurred in stochastic systems with bimodal or fat-tailed distributions that explore deterministically unstable regions.

\section*{Results}
\vspace{-1.5ex}
\subsection*{Definition of class of gene expression models}
\vspace{-1.5ex}
We consider the following class of models of gene expression control in which 
mature mRNA molecules $x$ and proteins $y$ of a gene of interest are synthesized and degraded stochastically in the following reactions \begin{equation}
\label{EQ: StochSys Def}
\begin{aligned}
\underbrace{x\myextralongrightarrow{f\left(\vec{\alpha};x,y,\vec{u}\right)} x+1}_{\text{mRNA synthesis}} && \underbrace{y \myextralongrightarrow{\gamma x} y+1}_{\text{Translation}} \phantom{.} \\
    \underbrace{x \myextralongrightarrow{\beta_x x} x-1}_{\text{mRNA Degradation}} && \underbrace{y \myextralongrightarrow{\beta_y y}y-1}_{\text{Protein Degradation}} .
\end{aligned}
\end{equation}
with parameters $\vec{\alpha},\beta_x,\beta_y,$ and $\gamma$.

\begin{figure}[b]
    \centering
    \vspace{-1ex}
\includegraphics{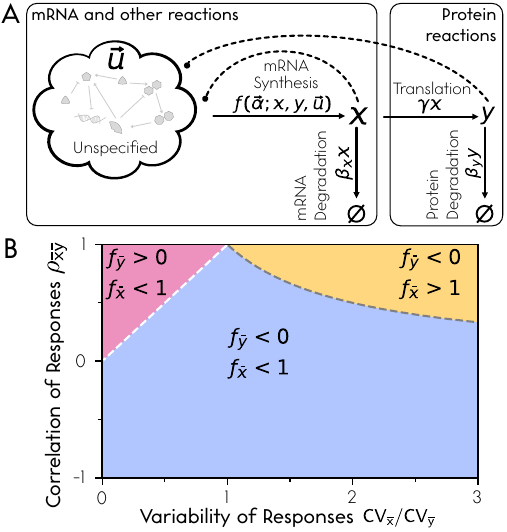}
\caption{\textbf{In the deterministic limit the response of gene expression models to random perturbations exhibits three distinct feedback regions.} (A) Illustration of the class of gene expression models defined in Eq.~\ref{EQ: StochSys Def}. (B) Infinitesimal random perturbations to rate parameters lead to three types of responses in the deterministic limit of mRNA and protein steady-states $\bar{x},\bar{y}$ under the assumption that perturbations affect the control network in a way such that direct effects to the protein dynamics module are statistically independent from effects on the rest.
In this regime the variability and correlation of the steady-state averages $\bar{x},\bar{y}$ across a set of random perturbations can thus be used to determine the nature of feedback sensitivities of gene expression regulation.
The sensitivity $f_{\bar{y}}$ quantifies the feedback response mediated through protein levels while $f_{\bar{x}}$ quantifies the feedback response that bypasses protein levels. The white and gray dashed lines respectively denote the boundary cases $f_{\bar{y}}=0$ and $f_{\bar{x}}=1$. }
\label{FIG:introfig}
\end{figure}

Here we assume the translation rate is proportional to mRNA levels and both species undergo first order degradation. 
However, the mRNA synthesis rate is left unspecified and allowed to depend in arbitrary ways on mRNA levels, protein levels, and a set of unspecified cellular components $\vec{u}$. The dynamics of these unspecified components are also allowed to depend on mRNA and protein levels in arbitrary ways. Therefore the reactions of Eq.~\ref{EQ: StochSys Def} do not define a specific model, but rather an entire class of models; $x$ and $y$ are two components of interest but are not assumed to be the only components in the cell. 
This generality captures complex cellular dynamics; for example making a mature mRNA molecule includes many steps including transcription and post-transcriptional processing~\cite{perez2013you}. Here the stochastic reaction in Eq.~\ref{EQ: StochSys Def} corresponds to the final step in this process. Together protein reactions form one module and reactions of mRNA and all other cellular components form another, see Fig.~\hyperref[FIG:introfig]{\ref*{FIG:introfig}A}.

In the deterministic limit, in which we ignore the inherent stochastic nature of the above biochemical reactions, the steady-state abundances $\overline{x}$ and $\overline{y}$ satisfy 
\begin{equation}
\label{EQ: DetermApprox}
        \overline{x} =\frac{1}{\beta_x} f\left(\vec{\alpha};\overline{x},\overline{y},\vec{\overline{u}}\big(\vec{k};\overline{x},\overline{y}\big)\right)\quad \mathrm{and}\quad
    \overline{y} = \frac{\gamma}{\beta_y} \overline{x}\quad.
\end{equation}
Here the steady-state abundances of the unspecified components are expressed in terms of their dependence on the mRNA and protein steady-state 
$\vec{\overline{u}} = \vec{\overline{u}}(\vec{k};\overline{x},\overline{y})$, with $\vec{k}$ denoting parameters in reactions of the unspecified components. 
The formulation of Eq.~\ref{EQ: DetermApprox} applies the implicit function theorem to a linear approximation of the unknown function, as is done in standard deterministic modular response analysis (MRA)~\cite{kholodenko2002untangling,borg2023modular}.

\vspace{-1ex}
\subsection*{Deterministic feedback constraints on the variability of responses to infinitesimal random perturbations}
\vspace{-1ex}
A set of infinitesimal random perturbations leads to a set of steady-states for mRNA and protein levels $\{(\bar{x}_0,\bar{y}_0),(\bar{x}_1,\bar{y}_1),(\bar{x}_2,\bar{y}_2),\ldots\}$ including the unperturbed levels $(\bar{x}_0,\bar{y}_0)$. 
This set of responses to random perturbations can be summarized in terms of their correlation coefficient and coefficients of variation 
\begin{equation}
    \label{EQ: Correlation Definition}
    \rho_{\bar{x}\bar{y}} \equiv \frac{\Cov\left[\bar{x},\bar{y}\right]}{\sqrt{\Var\left[\bar{x}\right]\Var\left[\bar{y}\right]}}\quad,\quad
    \textrm{CV}_\mathrm{\bar{x}} \equiv \frac{\sqrt{\Var\left[\bar{x}\right]}}{\textrm{E}\left[\bar{x}\right]} 
\end{equation}
and analogously for $\textrm{CV}_\mathrm{\bar{y}}$.

To analyze the response of the unperturbed steady state to random perturbations we introduce the feedback sensitivities
\begin{equation}
    \label{EQ: SensDef}
    \begin{aligned}
            f_{\bar{x}} \equiv \left. \frac{\partial \log f}{\partial\log \overline{x}} \right\rvert_{\bar{x},\bar{y}=\bar{x}_0,\bar{y}_0} &&
            f_{\bar{y}} \equiv \left. \frac{\partial \log f}{\partial\log \overline{y}} \right\rvert_{\bar{x},\bar{y}=\bar{x}_0,\bar{y}_0},
    \end{aligned}
\end{equation}
where the derivatives are of the function $f$ with the steady-state abundances of the unspecified components expressed in terms of their dependence on mRNA and protein steady-states as defined in Eq.~\ref{EQ: DetermApprox}, see Appendix~\ref{APP: ParamVarInfo}. The sensitivity coefficient $f_{\bar{y}}$ quantifies the strength of gene regulation feedback going through the protein, and $f_{\bar{x}}$ quantifies feedback from mRNA onto the mRNA synthesis rate that bypasses the gene-product levels. Such feedback may be relevant in processes such as gene interactions on the mRNA level through ribosomal competition~\cite{mather2013translational} or feedback on mature transcript synthesis from decapping and recapping processes~\cite{berry2022mrna,Ignatochkina2015}.

We consider infinitesimal random perturbations that affect multiple parameters simultaneously but from two statistically independent distributions for the two modules illustrated in Fig.~\hyperref[FIG:introfig]{\ref*{FIG:introfig}A}. This divides parameters into those that perturb the protein reactions directly ($\gamma$,$\beta_y$) from those that affect other cellular reactions $(\vec{\alpha},\beta_x,\vec{k})$. 
 Linearizing Eq.~\ref{EQ: DetermApprox} and imposing the statistical independence condition leads to (see Appendix~\ref{APP: ParamVarInfo})
\begin{equation}
\label{eq: sensitivity eq}
    \frac{f_{\bar{y}}}{1-f_{\bar{x}}} = \left(\frac{\textrm{CV}_{\bar{x}}}{\textrm{CV}_{\bar{y}}}-\rho_{\bar{x}\bar{y}}\right)/\left(\rho_{\bar{x}\bar{y}}-\frac{\textrm{CV}_{\bar{y}}}{\textrm{CV}_{\bar{x}}}\right).
\end{equation} 
Combining Eq.~\ref{eq: sensitivity eq} with the stability condition for the deterministic steady-states
\begin{equation}
    \label{eq: stability condition}
f_{\bar{x}} + f_{\bar{y}} < 1,
\end{equation}
then constrains the feedback sensitivities to three distinct regions as illustrated in Fig.~\hyperref[FIG:introfig]{\ref*{FIG:introfig}B}.
Where a measured system response lands can thus be used to determine
the sign of the protein feedback $f_{\bar{y}}$ and whether mRNA feedback control would be stable in the absence of protein feedback, i.e, whether $f_{\bar{x}}$ is less than 1 or not.

Like MRA, the above analysis of random perturbations is only exact for infinitesimal perturbations of deterministic systems. But in contrast to a standard application of MRA we do not assume the mRNA module has a sensitivity of $f_{\bar{x}} = -1$~\cite{kholodenko2002untangling}. Additionally, we allow for perturbations to directly affect multiple modules as long as those effects are statistically independent across modules.
\begin{figure}[b]
\centering
    \vspace{-0.5em}
\includegraphics{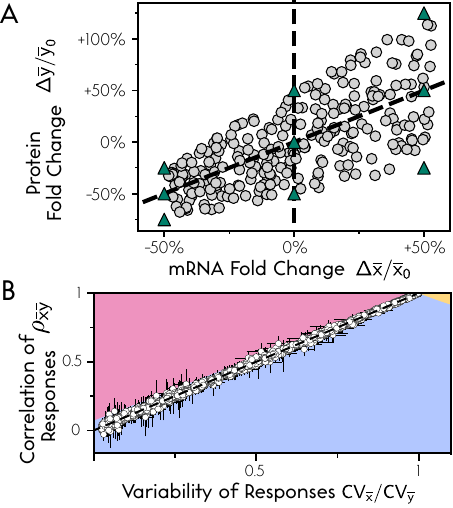}
\vspace{-1.5ex}
\caption{\textbf{Random perturbations of systems without protein feedback.}
(A) Example set of simulated average abundances of a stochastic system subject to random finite perturbations drawn from a statistically independent set of rate constant perturbations to protein production (translation) and mRNA synthesis. Dashed lines indicate the previously considered case of single-parameter perturbations to mRNA synthesis (diagonal) or translation rates (vertical). Finite sampling of random perturbations leads to noisy estimates for the correlation of responses. Throughout this work we exactly determined the large sample limit by computing responses to nine parameter sets (green triangles) as defined by the Cartesian product in Eq.~\ref{EQ: Cartesian Product of Perturbations}. (B) Numerically calculated responses of stochastic models of gene expression subject to statistically independent perturbations for systems without protein feedback (white dots). Error bars denote the 95\% confidence interval generated through a bootstrap distribution of averages based on three replicates per perturbation. In the absence of protein feedback the response variability of stochastic systems to finite perturbations is constrained to Eq.~\ref{EQ: FY0Constraint} (dashed line) predicted by deterministic approximations for infinitesimal perturbations. The presented sample includes systems with and without non-linear mRNA feedback~\cite{supp}.
    }
\vspace{-1.5ex}
    \label{fig:FiniteSample_NoFb}
\end{figure}

\vspace{-1.5ex}
\subsection*{Numerical simulations of statistically independent random perturbation experiments}
\vspace{-1.5ex}
Eqs.~\ref{eq: sensitivity eq} and~\ref{eq: stability condition} describe the response of deterministic systems to infinitesimal perturbations. However, real biological systems are stochastic and real experimental perturbations are finite. Previous work has considered the latter as a source of error in MRA and corrected for it using non-linear fitting techniques~\cite{borg2023modular,borg2024testing}. 

To account for the effect of stochastic fluctuations, we determined the average abundances of stochastic gene expression models within the class of Eq.~\ref{EQ: StochSys Def} through exact realizations~\cite{gillespie1977exact} for different parameter values. For simplicity, we primarily considered perturbations to the translation rate constant $\gamma$ and a multiplicative pre-factor $\lambda$ giving a mRNA synthesis rate of $\lambda f\left(\vec{\alpha};x,y,\vec{u},\right)$.
To determine the effect of non-multiplicative parameter perturbations, we also considered perturbation of $\lambda$ and $\gamma$ in systems with mRNA synthesis rates of the form $\lambda f\left(x,y\right) + c$.

Estimates of $\rho_{\bar{x}\bar{y}}$ and CV$_{\bar{x}}$, CV$_{\bar{y}}$ from a finite sample of perturbations will exhibit a sampling error. However, we can effectively determine the large sampling limit by constructing nine sets of statistically independent perturbations through the Cartesian product $ A\times B$ for the sets \begin{equation}
\begin{aligned}
\vspace{-1.5ex}
  A&= \{\lambda,\left(1-\Delta_\lambda\right)\lambda,\left(1+\Delta_\lambda\right)\lambda\}\\
  B&=\{\gamma,\left(1-\Delta_\gamma\right)\gamma,\left(1+\Delta_\gamma\right)\gamma\},
    \label{EQ: Cartesian Product of Perturbations}
\end{aligned} 
\end{equation}
see Fig.~\hyperref[fig:FiniteSample_NoFb]{\ref*{fig:FiniteSample_NoFb}A}. While a finite number of perturbations sampled from statistically independent distributions can exhibit significant sampling error, we find 100 random perturbations typically give accurate estimates~\cite{supp}.
\vspace{-3.5ex}
\subsubsection*{Responses of stochastic systems without protein feedback}
\vspace{-1.5ex}
If the mRNA synthesis rate does not (directly or indirectly) depend on protein levels mRNA levels are independent of $\gamma$ and $\beta_y$. Therefore under the statistical independence condition on parameters 
we have
$\textrm{E}\left[\overline{y}\right] = \textrm{E}\left[\gamma\right]
    \textrm{E}\left[\overline{x}\right]/\beta_y$ and
    $\textrm{E}\left[\overline{x}\overline{y}\right] = \textrm{E}\left[\gamma\right]\textrm{E}\left[\overline{x}^2\right]/\beta_y$
such that
\begin{equation}
    \Cov\left[\overline{x},\overline{y}\right] = \frac{\textrm{E}\left[\gamma\right]}{\beta_y} \Var[\overline{x}]\quad \implies \quad \label{EQ: FY0Constraint}
\rho_{\bar{x}\bar{y}} = \cvratiofrac.
\end{equation}

Eq.~\ref{EQ: FY0Constraint} is identical to the deterministic prediction for \mbox{$f_{\bar{y}}=0$} in the infinitesimal limit of Eq.~\ref{eq: sensitivity eq} and can thus be used to verify the validity of our numerical approach to analyze finite perturbations in non-linear stochastic systems. In particular, we considered a specific realization of Eq.~\ref{EQ: StochSys Def} with an mRNA synthesis rate of the form $\lambda h(x)$
for constant, positive, or negative Hill-functions $h(x)$.

Additionally we studied systems with an mRNA synthesis rate of the form $\lambda z$ set by a mediating molecule $z$ subject to stochastic reactions
\begin{equation}
\begin{aligned}
z \myextralongrightarrow{h(x)} z+1  &&
    z \myextralongrightarrow{\beta_z z} z-1,
\end{aligned}
\end{equation}
for various Hill-functions $h(x)$. 
The numerical data for the response variability satisfies the straight line prediction of Eq.~\ref{EQ: FY0Constraint} within expected confidence intervals, see Fig.~\hyperref[fig:FiniteSample_NoFb]{\ref*{fig:FiniteSample_NoFb}B}. See Supplementary Material~\cite{supp} for parameter values.

\vspace{-3.5ex}
\subsubsection*{Responses of stochastic systems with protein feedback}
\vspace{-1.5ex}
Next, we numerically determined the responses of example systems that include protein feedback with and without an intermediate molecule to 
test whether the observed response variability and correlations implied the same category of feedback sensitivity as the numerically calculated sensitivities defined in Eq.~\ref{EQ: SensDef}.

We constrained our perturbation sets such that perturbations did not change the qualitative feedback in order to ensure there is a ground truth feedback structure for the set of systems, such as positive protein feedback.
Our sample includes cases of both antagonistic and synergistic feedback combined via addition or multiplication. For a detailed description of systems and parameters, see Supplementary Material~\cite{supp}.

\begin{figure}[h]
    \centering
\includegraphics{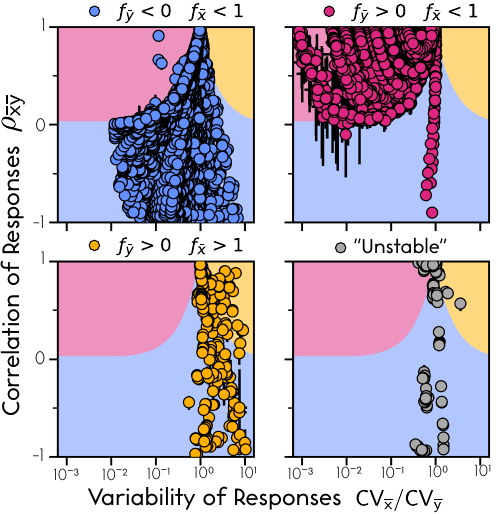} 
    \vspace{-1em}
    \caption{\textbf{Responses of stochastic systems to finite perturbations generally agree with deterministic constraints but exhibit non-trivial violations.}  Numerical responses of stochastic models of gene expression subject to statistically independent perturbations for a large sample of different feedback structures. Colored regions correspond to the deterministic predictions of Fig.~\ref{FIG:introfig}.
    Negative protein feedback systems with independently stable mRNA feedback dynamics (blue dots) generally agreed with deterministic bounds, but a small number of simulations violated the bounds.  Systems with positive protein feedback and mRNA feedback dynamics that are independently stable (red dots) showed general agreement with constraints but a subset with antagonistic protein feedback systems violated the bounds. Many systems in which mRNA feedback was unstable in the absence of a stabilizing protein feedback (yellow dots) violate the deterministic response variability constraints. Some simulated systems exhibited sensitivities at their average levels that would not be possible for deterministic systems because they violate the stability constraint of Eq.~\ref{eq: stability condition} (gray dots). Simulation data correspond to different feedback structures: feedback from only one of the molecules, synergistic feedback from both molecules and antagonistic feedback from protein or both molecules. We considered both additive and multiplicative feedback with multiple control inputs and feedback with and without intermediate molecules~\cite{supp}.}

\label{FIG:Alldots_atavg}
    \vspace{-1.5ex}
\end{figure}

Responses of systems with protein feedback levels generally obeyed the predicted deterministic constraints, see Fig.~\ref{FIG:Alldots_atavg}. This was true for positive and negative protein feedback in the presence and absence of mRNA feedback. However, all feedback categories contain at least one example of a system whose response non-trivially violates the bounds. 

A significant proportion of our observed violations occurred in systems with large and positive mRNA feedback.
However, this is a non-quantitative measure as it is biased by our system sample. Additionally some sensitivities of stochastic systems evaluated at their averages violated the stability condition Eq.~\ref{eq: stability condition}, highlighting that deterministic constraints do not straightforwardly apply to average abundances of stochastic systems.

\vspace{-1.5ex}
\subsubsection*{Average sensitivities \emph{vs.} sensitivities evaluated at the average}
\vspace{-1.5ex}
\begin{figure*}[htb]
    \centering
    \includegraphics[width=\textwidth]{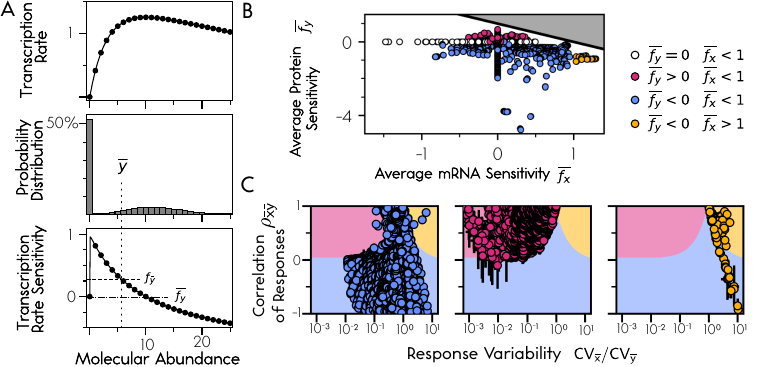}
    \caption{\textbf{Categorizing systems by their \emph{average} sensitivity removes violations of the stability condition but does not eliminate response constraint violations.} (A) Sensitivity measures of deterministic steady states do not uniquely generalize to stationary states of stochastic systems. For example, we can consider a stochastic system's feedback sensitivity evaluated at its distribution average, $f_{\bar{y}}$, or their mean sensitivity averaged over the stationary state distribution, $\bar{f_y}$. Example system with a non-monotonic rate function and switch like behavior to illustrate that in extreme cases $\bar{f_y}$ and $f_{\bar{y}}$ can differ qualitatively. (B) Average sensitivities of all sampled systems (dots) satisfied the deterministic stability condition of Eq.~\ref{eq: stability condition} (excluded gray region) in contrast to sensitivities evaluated at the average. (C) Using average sensitivity classifications some systems became consistent with deterministic constraints but others became inconsistent.}
\label{FIG:avg_sensitivity_res}
    \vspace{-1.5ex}
\end{figure*}
Under stochastic dynamics a molecule's average abundance can correspond to a rarely visited state,
which might explain the discrepancy between deterministic constraints and the behavior of stochastic systems. To capture effect of feedback sensitivities across all visited states we next considered the \emph{average} sensitivities
\begin{equation}
    \label{EQ: AvgSensDef}
    \begin{aligned}
            \overline{f_x} &\equiv \sum_{x,y} P\left(x,y\right) \frac{\partial \log f}{\partial\log x}\hspace{1ex}, \hspace{1ex}
            \overline{f_y} &\equiv \sum_{x,y} P\left(x,y\right) \frac{\partial \log f}{\partial\log y}
    \end{aligned}
\end{equation}
where the previously defined derivatives of Eq.~\ref{EQ: SensDef} are averaged over the observed stationary state distribution $P(x,y)$.
As before, we restricted our consideration to systems that did not change sign in either sensitivity across the perturbations.

No system violated the stability constraint of Eq.~\ref{eq: stability condition} when sensitivities were defined through Eq.~\ref{EQ: AvgSensDef}, suggesting that average sensitivities more meaningfully characterize stochastic systems compared to sensitivities evaluated at the average, see Fig.~\hyperref[FIG:avg_sensitivity_res]{\ref*{FIG:avg_sensitivity_res}B}.
Additionally, 
significantly fewer systems qualitatively violated the deterministic constraints. 
However, these violations were non-trivial and demonstrated that average sensitivities do not ensure full consistency with deterministic constraints either, see Fig.~\hyperref[FIG:avg_sensitivity_res]{\ref*{FIG:avg_sensitivity_res}C}. Since some systems obeyed bounds when classified by one sensitivity measure and not the other, neither average sensitivities nor sensitivities at the average can be considered fundamentally more accurate in describing response variability through a deterministic linear response model.

\vspace{-3.5ex}
\subsubsection*{Linear response of stochastic systems} 
\vspace{-1.5ex}
The data in Figs.~\ref{FIG:Alldots_atavg},~\ref{FIG:avg_sensitivity_res} characterize the response of non-linear stochastic systems to finite perturbations of various magnitudes.
Because large perturbations will affect the accuracy of linear response predictions for nonlinear systems even in deterministic systems~\cite{borg2023modular}, 
we next focus on small perturbations of stochastic systems to explore the regime in which the response of average abundances are linear.

\begin{figure*}[htb]
    \centering
\includegraphics[width=\textwidth]{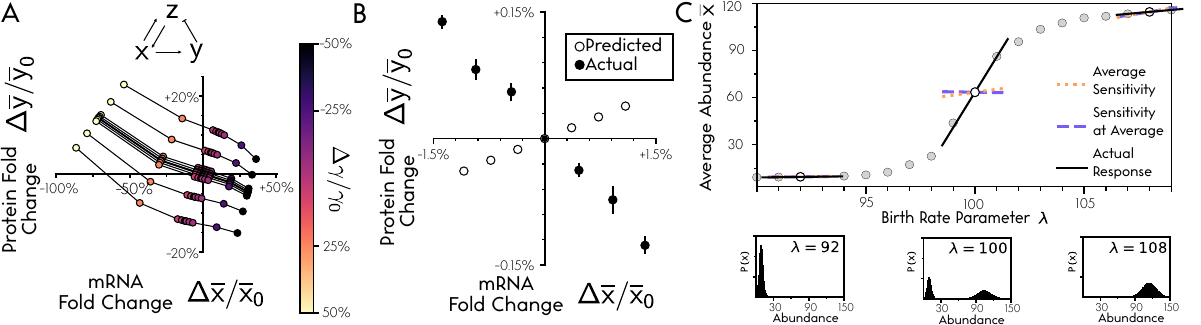}
      \vspace{-2em}
    \caption{\textbf{Deterministic predictions can fail to predict even the linear response of stochastic systems.}
(A) Non-linear example response of a feedback system with an intermediate component that positively affects the mRNA synthesis rate but is itself regulated antagonistically through a negative feedback from protein levels and a positive feedback from mRNA levels, as defined in Appendix~\ref{APP: StochViol_System}.  (B) For very small perturbations (0.5\%,1\%, 1.5\% to $\gamma$) the response is approximately linear (black dots) with a slope of opposite sign compared to that predicted by the deterministic limit using average sensitivity (white dots). Error bars denote the standard error across three replicate simulations.  
(C) Average abundances (gray dots) of a one-dimensional system with positive feedback in its production given by Eq.~\ref{EQ: OneDSysDef} with rate constants $\beta=1,c=0.1,K=50,n=4$. The linear response coefficient of the stochastic system (solid tangent lines) differs drastically from deterministic sensitivity analyses (dashed lines) when the probability distribution is bimodal. 
Both deterministic response predictions are accurate approximations when the stochastic stationary state distribution is unimodal.
}
\label{FIG: Stochastic Linear Response}
   \vspace{-1em}
\end{figure*}

In particular, we consider an example realization of Eq.~\ref{EQ: StochSys Def} in which mRNA synthesis is controlled by an intermediate species whose own production rate is formed by antagonistic additive feedback combining positive mRNA feedback and negative protein feedback, see Fig.~\ref{FIG: Stochastic Linear Response}A and Appendix \ref{APP: StochViol_System} for details. Its numerical average sensitivities (and sensitivities at the average) correspond to the yellow region for systems in which the global mRNA feedback is stabilized by the protein feedback. 
As expected, the response of average mRNA and protein abundances is non-linear over large perturbations (Fig.~\hyperref[FIG: Stochastic Linear Response]{\ref*{FIG: Stochastic Linear Response}A}). However, the discrepancy between the deterministic prediction and stochastic system responses persists under perturbations as small as 0.1\% and 1\% for which the response is approximately linear (Fig.~\hyperref[FIG: Stochastic Linear Response]{\ref*{FIG: Stochastic Linear Response}B}). 
In this example system deterministic theory predicts a positive slope for mRNA-protein fold changes but the actual slope is negative. As a result the category of feedback in the system would be misidentified by modular response analysis in a manner not attributable to the previous analyses of measurement noise~\cite{Andrec2005Inference,thomaseth2018impact} or the finite size of perturbations~\cite{borg2023modular,borg2024testing}.

The discrepancy between the response of stochastic systems and deterministic predictions in the linear response regime can be intuitively understood considering the following one-dimensional positive feedback system
\begin{equation}
\label{EQ: OneDSysDef}
\begin{aligned}
    x \myextralongrightarrow{\lambda \left(c+\frac{x^n}{x^n+K^n}\right)} x+1  &&
    x \myextralongrightarrow{\beta x} x-1.
\end{aligned}
\end{equation}
This systems exhibits a stochastic phase transition between high and low abundance modes,
see Figure~\ref{FIG: Stochastic Linear Response}C.
In the high and low abundance regimes, the linear response coefficient based on the sensitivity at the average and the average sensitivity agree and accurately predict the response of the stochastic system. However, during the phase transition both sensitivity-based estimates  differ drastically from the response of the stochastic system to infinitesimal perturbations.
\begin{figure}[b]
    \centering
        \vspace{-2em}
\includegraphics{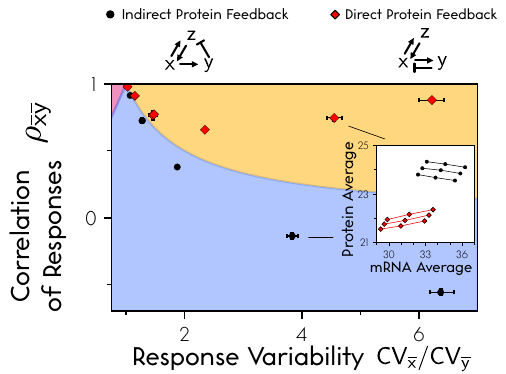}
    \vspace{-2em}
    \caption{\textbf{Feedback systems with equivalent deterministic steady state dynamics can exhibit significantly different responses in the stochastic regime.}
Responses of two example systems with identical steady-states and sensitivities in the deterministic regime.  In one system protein (black dots) feedback indirectly affects the mRNA synthesis rate through an intermediate as defined in Fig.~\ref{FIG: Stochastic Linear Response}A while in the other system (red diamonds) protein feedback directly controls the mRNA synthesis rate, see Appendix~\ref{APP: StochViol_System}. Error bars denote the 95\% confidence interval. Inset: Average abundances exhibit approximately linear responses~\cite{supp} but with slopes of opposite signs.}
\label{FIG: DirectIndirect}
\end{figure}

This simple one-dimensional example illustrates how the observed linear response of a stochastic system is shaped by its entire probability distribution which cannot necessarily predicted by local rate sensitivities alone. 
While previous work has noted that MRA breaks down for positive feedback systems with a steady state discontinuity~\cite{thomaseth2018impact}, the above system does not exhibit a discontinuity in its average abundance
because stochastic effects wash out the discontinuity of its deterministic analogue.

To further illustrate how the stochastic dynamics of biochemical systems fundamentally affect their responses to perturbations, we compared the example system of Fig.~\ref{FIG: Stochastic Linear Response}A with a system in which the protein feedback is direct rather than through an intermediate, see Fig.~\ref{FIG: DirectIndirect} and Appendix \ref{APP: StochViol_System}. The two systems have identical noise-free approximations for mRNA and protein steady-states and thus identical predicted linear responses in the deterministic regime. However, the two stochastic realizations of the systems exhibit qualitatively different responses, see Fig.~\ref{FIG: DirectIndirect}. While the deterministic approximations of the systems are identical, the stochastic dynamics leads to differences in the averages, mode, and skew of the stationary states~\cite{supp}.

\section*{Discussion}
\vspace{-2ex}
The control of gene expression is at the heart of cellular life. Here we present results for a broad class of feedback control models of gene expression to illustrate how existing modular response approaches like MRA can be generalized to random perturbations that affect multiple modules simultaneously.
Our results extend the applicability of the MRA approach from single perturbations with a known modular effect to a set of perturbations that hit two modules in a statistically independent fashion. Even when statistical independence cannot be strictly guaranteed this approach may still be a reasonable approximation to interpret high-throughput experiments with hundreds or thousands of drug perturbations. While our class of gene expression models with two modules is a particular case study of the MRA approach, the response of more complex modular structures to random perturbations can be determined similarly.

Accounting for the fact that cellular processes are stochastic rather than deterministic, we showed how predictions from a deterministic framework generally agreed with the numerical behavior of stochastic test cases. However, our results also reiterate that deterministic analyses fundamentally cannot be guaranteed to predict the average response of stochastic systems---even in the limit of infinitesimal perturbations~\cite{van1971case}. That is because in stochastic systems a linearized analysis necessarily approximates rates 
over the entire range of stochastically visited states. This is a much more severe approximation than might be intuitively expected when working within a deterministic framework, where linear approximations need only be valid over the range of the perturbation. This difference explains why example systems with bimodal distributions or fat tails did not agree with the deterministic predictions in our analysis. Unrelated complications, such as experimental coarse graining of chemical species, or unaccounted perturbation effects across multiple modules, can lead to contradictory answers even in deterministic modular response analyses~\cite{prabakaran2014paradoxical}. However, our reported discrepancies are entirely due to the inherently stochastic dynamics of biochemical reactions in cells and are not related to such additional complications.

The fact that stochastic variability affects the response of populations to perturbations is biologically relevant in many contexts. For example, transient bacterial subpopulations survive antibiotic treatment that would kill the population at its average~\cite{harms2016mechanisms} or cell fates during embryonic development are the result of the differentiation of subpopulations~\cite{goolam2016heterogeneity}. A limitation of our study is that our numerical example systems are not guaranteed to be representative of biological systems. Further work remains to be done to establish general analytical results that connect the response of average abundances in stochastic biochemical feedback systems to their underlying biochemical rate functions.

\section*{Author Contributions}
SI derived the analytical results and performed the numerical simulations. SI and AH wrote the article.

\section*{Declaration of Interests}
The authors declare that they do not have any competing interests.

\section*{Acknowledgments}
This work took place on the treaty lands of the Mississaugas of the Credit, the unceded lands of the Wolastoqiyik, and on the traditional lands of the Seneca and the Wendat. We thank B Kell, Raymond Fan, Euan Joly-Smith, Linan Shi, and Johnathan Sorrentino for helpful discussions and suggestions. Computational facilities for this work were provided by the UTM High Performance Computing Cluster. This work was supported by the Natural Sciences and Engineering Research Council of Canada. SI gratefully acknowledges financial support from the Arthur and Sandra Irving Foundation's C.~David Naylor Fellowship, the Ontario Graduate Scholarship, the Walter C.~Sumner Memorial Fellowship, and the Vanier Canada Graduate Scholarship. 

\section*{Appendix}
\renewcommand{\thesubsection}{\Alph{subsection}}
\subsection{Deterministic feedback constraints for infinitesimal random perturbations}
\label{APP: ParamVarInfo}
To determine the sensitivities of the molecular averages $\overline{x},\overline{y}$ for the system defined in Eq.~\ref{EQ: DetermApprox} we distinguish between parameters that \emph{directly} affect protein reactions
\begin{equation}
    P= \{\beta_y,\gamma\}
\end{equation}
and ``mRNA module'' parameters that \emph{directly} affect any of the remaining reactions, given by the set
\begin{equation}
    M \in \{\beta_x,\alpha_1,...\alpha_j,k_1,...k_{q}\}.
\end{equation}

We expand Eq.~\ref{EQ: DetermApprox} to first order around the average value of parameters across a perturbation experiment---denoted by ${m_i}_0$ and ${p_i}_0$---and the corresponding mRNA and protein levels $\overline{x}_0$ and $\overline{y}_0$. To first order, the fold-change in the mRNA and protein levels is then given by
\begin{equation}
\label{eq: TaylorExpand}
    \begin{aligned}
        \frac{\Delta \overline{x}}{\overline{x}_0} &= \sum_{m_i\in M}\loglog{\overline{x}}{m_i}\frac{\Delta m_i}{{m_i}_0} +\sum_{m_i\in M}\loglog{\overline{x}}{p_i}\frac{\Delta p_i}{{p_i}_0} \\ 
        \frac{\Delta \overline{y}}{\overline{y}_0} &= \sum_{p_i\in P}\loglog{\overline{y}}{m_i}\frac{\Delta m_i}{{m_i}_0} +\sum_{p_i \in P}\loglog{\overline{y}}{p_i}\frac{\Delta p_i}{{p_i}_0}.
    \end{aligned}
\end{equation}
In the infinitesimal perturbation limit, the average parameter values are equivalent to the unperturbed parameters and the molecular levels represent the unperturbed state.

By treating the fold-changes of the parameters in Eq.~\ref{eq: TaylorExpand} as random variables we calculate the covariance of the molecular abundances. Combining the statistical independence assumption between any mRNA and protein module parameters, which implies $\Cov\left[m_i,p_j\right] = 0$, and the bilinearity of the covariance operator gives
\begin{equation}
\label{APEQ: Covariance of Mols}
\begin{split}
        \Cov\left[\frac{\Delta \overline{x}}{\overline{x}_0},\frac{\Delta \overline{y}}{\overline{y}_0}\right] = 
        {\sum_{p_i,p_j \in P}}\hspace{-1.5ex}\Cov\left[\loglog{\overline{x}}{p_i}\frac{\Delta p_i}{p_{i0}},\loglog{\overline{y}}{p_j}\frac{\Delta p_j}{p_{j0}}\right]\\
        \hspace{1em}+{\sum_{m_i,m_j \in M}}
        \hspace{-1.5ex}\Cov\left[\loglog{\overline{x}}{m_i}\frac{\Delta m_i}{m_{i0}},\loglog{\overline{y}}{m_j}\frac{\Delta m_j}{m_{j0}}\right].
    \end{split}
\end{equation}
Introducing the normalized covariance of two random variables $u,v$
\begin{equation}
    \eta_{uv} \equiv \frac{\Cov\left[u,v\right]}{\bar{u}\bar{v}},
\end{equation}
we can eliminate the perturbation size dependence in Eq.~\ref{APEQ: Covariance of Mols}
\begin{multline}
    \eta_{\bar{x}\bar{y}} = {\sum_{m_i,m_j \in M}} \loglog{\overline
    {x}}{m_i}\loglog{\overline{y}}{m_j}\eta_{m_im_j} \\ + {\sum_{p_i,p_j \in P}} \loglog{\overline
    {x}}{p_i}\loglog{\overline{y}}{p_j}\eta_{p_ip_j}.
    \label{APEQ: etaXY in terms of etaPARAs}
\end{multline}

To relate Eq.~\ref{APEQ: etaXY in terms of etaPARAs} to the sensitivities defined in Eq.~\ref{EQ: SensDef} we  distinguish between dependencies on the specified components $\overline{x},\overline{y}$ and the unspecified components $\overline{u}_i$
\begin{widetext}
\begin{equation}
    \begin{aligned}
        f_{\overline{y}} &\equiv \left(\loglog{f}{\overline{y}}\right)_{\vec{\overline{u}},\overline{x}}+\sum_{i}\left(\loglog{f}{\overline{u}_i}\right)_{\overline{u}_{j\neq i},\overline{x},\overline{y}}\left(\loglog{\overline{u}_i}{\overline{y}}\right)_{\overline{u}_{j\neq i},\overline{x}} \\
        f_{\overline{x}} &\equiv \left(\loglog{f}{\overline{x}}\right)_{\vec{\overline{u}},\overline{y}}+\sum_{i}\left(\loglog{f}{\overline{u}}\right)_{\overline{u}_{j\neq i},\overline{x},\overline{y}}\left(\loglog{\overline{u}_i}{\overline{x}}\right)_{\overline{u}_{j\neq i},\overline{y}},
    \end{aligned}
\end{equation}
\end{widetext}
where the notation
\begin{equation}
    \left(\loglog{f}{t}\right)_{a,b,\ldots}
\end{equation}
refers to the partial derivative of the function $f$ with respect to $t$ keeping $a,b$ etc. constant. Parameters are always assumed to be fixed when the derivative is with respect to $\overline{x}$, $\overline{y}$, $\vec{\overline{u}}$, or other parameters.

Through implicit differentiation of the deterministic equation for $\overline{y}$ in Eq.~\ref{EQ: DetermApprox}, we can write the protein abundance sensitivities in terms of the mRNA abundance sensitivities
\begin{equation}
    \label{eq: implicitdiff_1}
    \begin{aligned}
        \loglog{\overline{y}}{m_i} &= &&\loglog{\overline{x}}{m_i} \\
        \loglog{\overline{y}}{\gamma} &= &&1 +\loglog{\overline{x}}{\gamma} \\
        \loglog{\overline{y}}{\beta_y} &= &&\loglog{\overline{x}}{\beta_y} - 1.
    \end{aligned}
\end{equation}
Introducing
\begin{equation}
    S \equiv 1 - f_{\overline{x}} - f_{\overline{y}},
\end{equation}
we use implicit differentiation of Eq.~\ref{EQ: DetermApprox} combined with Eq.~\ref{eq: implicitdiff_1} to derive the sensitivity of mRNA levels to mRNA module parameters
\begin{equation}
        \begin{aligned}
        \loglog{\overline{x}}{\beta_x} &= -\frac{1}{S}  \\
        \loglog{\overline{x}}{\alpha_i} &= \frac{1}{S}\left(\loglog{f}{\alpha_i}\right)_{\vec{\overline{u}},x,y} \\
        \loglog{\overline{x}}{k_i} &= \frac{1}{S}\left(\sum_{\ell}\left(\loglog{f}{\overline{u}_\ell}\right)_{\overline{u}_{n\neq \ell},\bar{x},\bar{y}}\left(\loglog{\overline{u}_\ell}{k_i}\right)_{\overline{u}_{n\neq \ell},\bar{x},\bar{y}}\right).
    \end{aligned}
\end{equation}
Similarly, we obtain for the sensitivities of mRNA abundances to protein parameters
\begin{equation}
\label{eq: mrna_to_proteinparams}
    \begin{aligned}
        \loglog{\overline{x}}{\gamma} &= \frac{f_{\overline{y}}}{S}  \\
        \loglog{\overline{x}}{\beta_y} &= -\frac{f_{\overline{y}}}{S}.
    \end{aligned}
\end{equation}
The protein sensitivities to mRNA module perturbations follow from Eq.~\ref{eq: implicitdiff_1}. For the protein parameters substituting Eq.~\ref{eq: mrna_to_proteinparams} into Eq.~\ref{eq: implicitdiff_1} yields
\begin{equation}
    \begin{aligned}
        \loglog{\overline{y}}{\gamma} &= \frac{1-f_{\overline{x}}}{S}\\
        \loglog{\overline{y}}{\beta_y} &= \frac{f_{\overline{x}}-1}{S}. \\
    \end{aligned}
\end{equation}

Defining the parameter variability measures \begin{widetext}
    \begin{equation}
    \begin{split}
    N_p &\equiv \eta_{\gamma\gamma} + \eta_{\beta_y\beta_y} - 2\eta_{\gamma\beta_y} \\
N_m &\equiv \eta_{\beta_x\beta_x} + 
\sum_i{\sum_j{\left(\loglog{f}{\alpha_i}\right)_{\vec{\overline{u}},x,y}\left(\loglog{f}{\alpha_j}\right)_{\vec{\overline{u}},x,y}\eta_{\alpha_i\alpha_j}}} \\&~~+
\sum_i{\sum_j{\left(\sum_{\ell}\left(\loglog{f}{\overline{u}_\ell}\right)_{\overline{u}_{n\neq \ell},\bar{x},\bar{y}}\left(\loglog{\overline{u}_\ell}{k_i}\right)_{\overline{u}_{n\neq \ell},\bar{x},\bar{y}}\right)\left(\sum_{s}\left(\loglog{f}{\overline{u}_s}\right)_{\overline{u}_{n\neq s},\bar{x},\bar{y}}\left(\loglog{\overline{u}_s}{k_j}\right)_{\overline{u}_{n\neq \ell},\bar{x},\bar{y}}\right)   \eta_{k_ik_j}}}\\&~~+
2\sum_i\sum_j\left(\loglog{f}{\alpha_i}\right)_{\vec{\overline{u}},x,y}\left(\sum_{\ell}\left(\loglog{f}{\overline{u}_\ell}\right)_{\overline{u}_{n\neq \ell},\bar{x},\bar{y}}\left(\loglog{\overline{u}_\ell}{k_j}\right)_{\overline{u}_{n\neq \ell},\bar{x},\bar{y}}\right)\eta_{\alpha_ik_j}\\&~~-
2\left(\sum_i{\left(\loglog{f}{\alpha_i}\right)_{\vec{\overline{u}},x,y}}\eta_{\beta_x\alpha_i}+\sum_i\left(\sum_{\ell}\left(\loglog{f}{\overline{u}_\ell}\right)_{\overline{u}_{n\neq \ell},\bar{x},\bar{y}}\left(\loglog{\overline{u}_\ell}{k_i}\right)_{\overline{u}_{n\neq \ell},\bar{x},\bar{y}}\right)\eta_{\beta_xk_i}\right),
    \end{split}
     \end{equation}
\end{widetext}
and substituting them into Eq.~\ref{APEQ: etaXY in terms of etaPARAs} then gives the covariance of molecular abundances
\begin{equation}
\label{eq: detcovariance}
    \eta_{\bar{x}\bar{y}} = \left(\frac{1}{S}\right)^2\left(N_m + \left(1-f_{\overline{x}}\right)f_{\overline{y}} N_p\right).
\end{equation}
Following the same process for the variance of the mRNA and protein levels we find
\begin{equation}
\label{eq: detvariability}
    \begin{aligned}
        \eta_{\bar{x}\bar{x}} &= \left(\frac{1}{S}\right)^2\left(N_{m}+f_{\bar{y}}^2N_p\right) \\
        \eta_{\bar{y}\bar{y}} &= \left(\frac{1}{S}\right)^2\left(N_{m}+\left(1-f_{\bar{x}}\right)^2N_p\right).
    \end{aligned}
\end{equation}
Eq.~\ref{eq: sensitivity eq} follows by substitution of Eqs.~\ref{eq: detcovariance} and~\ref{eq: detvariability} into Eq.~\ref{EQ: Correlation Definition}.

\subsection{Indirect and direct feedback systems definition}
\label{APP: StochViol_System}
We define the following two Hill functions
\begin{equation}
    \begin{aligned}
        h^+\left(x\right) &= \frac{x^6}{x^6+40^6} \quad,\quad
        h^-\left(y\right) &= \frac{20^5}{y^5+20^5}
    \end{aligned}
\end{equation}
which represent a positive feedback from mRNA and negative feedback from protein respectively. We defined two systems with an additional ``unobserved'' intermediate component $z$ based on the form in Eq.~\ref{EQ: StochSys Def} and parameters $\lambda=4.5, \beta_x=0.1, \gamma = 0.7, \beta_y = 1$. 

The \emph{indirect} feedback system analyzed in Fig.~\ref{FIG: Stochastic Linear Response}A,B and \ref{FIG: DirectIndirect} had an mRNA synthesis rate $    f\left(z\right) = \lambda z$
with an intermediate component whose dynamics follows
\begin{equation}
\begin{aligned}
    z \myextralongrightarrow{h^+\left(x\right)+h^-\left(y\right)} z+1  &&
    z \myextralongrightarrow{z}z-1
\end{aligned}
\end{equation}
which represents a system where the feedback from mRNA and protein move through the same intermediate, whose abundance sets the mRNA synthesis rate.

The data in Fig.~\ref{FIG: Stochastic Linear Response}B, corresponds to the above system, perturbing $\gamma$ by 0.5\%,1\%, and 1.5\%\, respectively, 
while holding $\lambda$ constant. The white dots correspond to the linear prediction based on the average sensitivity for the deviation of those points. Perturbations included increases and decreases by the above fold-change values. In the inset, we see the results of a single example perturbation experiment where $\Delta\lambda/\lambda_0 = 0.01$ and $\Delta\gamma/\gamma_0 = 0.05$. 

The corresponding \emph{direct} protein feedback system shown in Figure \ref{FIG: DirectIndirect} has identical parameters, but instead has a mRNA synthesis rate $f\left(y,z\right) = \lambda\left(z+h^-\left(y\right)\right)$
and intermediate component dynamics of
\begin{equation}
\begin{aligned}
    z \myextralongrightarrow{h^+\left(x\right)} z+1  &&
    z \myextralongrightarrow{z}z-1
\end{aligned}
\end{equation}
where the abundance of $y$ influences the mRNA synthesis with the same functional form, but affects the $x$ production rate and not the intermediate component's production rate.

The degradation rate constant for $z$ has been set to unity which leads to the above direct and indirect feedback systems exhibiting identical noise-free approximations for the average levels in $x$, given by
\begin{equation}
    \overline{x} = \frac{\lambda}{\beta_x}\left(h^+\left(\overline{x}\right)+h^-\left(\overline{y}\right)\right).
\end{equation}

\bibliography{bibliography}
\end{document}